# The science case for POLLUX, a high-resolution UV spectropolarimeter onboard LUVOIR

Jean-Claude Bouret*[a], Coralie Neiner[b], Ana I. Gómez de Castro[c], Chris Evans[d], Boris Gaensicke[e], Steve Shore[f], Luca Fossati[g], Cécile Gry[a], Stéphane Charlot[h], Frédéric Marin[i], Pasquier Noterdaeme[h], Jean-Yves Chaufray[j]

[a]Aix Marseille Univ, CNRS, CNES, LAM, Marseille, France;
[b]LESIA, Observatoire de Paris, Université PSL, CNRS, Sorbonne Université, Univ. Paris Diderot, Sorbonne Paris Cité, 5 place Jules Janssen, 92195 Meudon, France;
[c]Universidad Complutense de Madrid, AEGORA Research Group, Fac. CC. Matematicas, Plaza de Ciencias 3, 28040 Madrid, Spain;
[d]UK Astronomy Technology Centre, Royal Observatory, Blackford Hill, Edinburgh, EH9 3HJ, UK;
[e]University of Warwick, Department of Physics, Hibbet Hill Road, Coventry CV4 7AL, UK;
[f]Dipartimento di Fisica "Enrico Fermi"Università di Pisa, 56127, Pisa, Italy
[g]Space Research Institute, Austrian Academy of Sciences, Schmiedlstrasse 6, 8042, Graz, Austria;
[h]Sorbonne Universités, UPMC-CNRS, UMR7095, Institut d'Astrophysique de Paris, F-75014, Paris, France;
[i]Université de Strasbourg, CNRS, Observatoire Astronomique de Strasbourg, UMR 7550, F-67000 Strasbourg, France;
[j]LATMOS/IPSL, UPMC Univ. Paris 06 Sorbonne Universités, UVSQ, CNRS, 75252 Paris Cedex 05, France;

**ABSTRACT**

POLLUX is a high-resolution, UV spectropolarimeter proposed for the 15-meter primary mirror option of LUVOIR[1]. The instrument Phase 0 study is supported by the French Space Agency (CNES) and performed by a consortium of European scientists. POLLUX has been designed to deliver high-resolution spectroscopy (R ≥ 120,000) over a broad spectral range (90-390 nm). Its unique spectropolarimetric capabilities will open-up a vast new parameter space, in particular in the unexplored UV domain and in a regime where high-resolution observations with current facilities in the visible domain are severely photon starved.

POLLUX will address a range of questions at the core of the LUVOIR Science portfolio. The combination of high resolution and broad coverage of the UV bandpass will resolve narrow UV emission and absorption lines originating in diffuse media, thus permitting the study of the baryon cycle over cosmic time: from galaxies forming stars out of interstellar gas and grains, and stars forming planets, to the various forms of feedback into the interstellar and intergalactic medium (ISM and IGM), and active galactic nuclei (AGN). UV circular and linear polarimetry will reveal the magnetic fields for a wide variety of objects for the first time, from AGN outflows to a diverse range of stars, stellar explosions (both supernovae and their remnants), the ISM and IGM. It will enable detection of polarized light reflected from exoplanets (or their circumplanetary material and moons), characterization of the magnetospheres of stars and planets (and their interactions), and measurements of the influence of magnetic fields at the (inter)galactic scale. In this paper, we outline the key science cases of POLLUX, together with its high-level technical requirements. The instrument design, its estimated performances, and the required technology development are presented in a separated paper[2].

*jean-claude.bouret@lam.fr; (+33) 491056902

# 1. INTRODUCTION

The major challenge of contemporary astrophysics is to advance our understanding of the origin and evolution of galaxies, stars and planets that make up our Universe, and the life within it. The Large Ultraviolet/Optical/Infrared Surveyor (LUVOIR) proposed as one of the four "flagship" mission concept studies led by NASA for the 2020 Decadal Survey is designed to address this challenge and the related science cases. A versatile suite of instruments is envisioned to shed new light on the fascinating questions at the core of the LUVOIR vision. Under the leadership of LAM and LESIA (France), European institutes have come together to propose an instrument that would be onboard the 15-meter primary mirror option of LUVOIR. This instrument, POLLUX, is a high-resolution ($R \geq 120,000$) spectropolarimeter, operating at UV wavelengths (90-400 nm). The Phase 0 study for POLLUX funded by CNES, started in January 2017.

The UV wavelength range is a treasure trove for astrophysics as it provides many of the most powerful probes of the hot and cold ISM and IGM gas, protostellar and protoplanetary disks, (exo)-planetary atmospheres, the peak emission from hot stellar atmospheres, star-forming galaxies, and AGN outflows. The major diagnostics for these objects are typically in the rest-frame UV between 100-200 nm and remain below the atmospheric cutoff (at 310 nm) for redshifts $z \lesssim$ 1-2. Futher, extreme highly-ionized species formed in hot, diffuse gas produce emission lines in the EUV range that can be redshifted into the near-UV domain. Only with a space-borne facility can we benefit from this unrivaled diagnostic power of the UV. POLLUX will observe over the spectral range from 90 to 390 nm, covering a wide range of species and temperatures that probe most astrophysical processes, for a large variety of objects and structures in the Universe.

POLLUX will deliver $R \geq 120,000$ throughout its spectral range, enabling resolution of the numerous and narrow UV emission and absorption lines (with FHWMs of a few km s$^{-1}$). This includes important molecules such as $H_2$ and CO, which are intrinsically far stronger than the molecular emission bands accessible in the near-IR (e.g. with the James Webb Space Telescope). Access to these molecular bands is required to infer the density profile of planet-forming disks, their temperature, kinematics, and mass-transfer rates. Similarly, POLLUX will identify the pattern of these molecules (plus possibly $H_2O$) in the upper atmosphere of (transiting) exoplanets, thereby revolutionizing our understanding of these objects. At such high spectral resolution, we can even imagine studies of winds on exoplanets if the atmospheric Doppler shift they induce differs from the shift due to the planet's orbital motion. Other topics for which high-resolution UV spectroscopy with POLLUX is essential include studies of the atmospheric dynamics of cool stars, their circumstellar medium and astrospheres within 100 pc around the Sun, as well as studies of the baryon cycle within galaxies, the ISM and the IGM, which requires resolving velocity structures at 2-5 km s$^{-1}$.

**The most innovative characteristic of POLLUX is its unique spectropolarimetric capability**. This will enable studies of a broad range of important astrophysical environments, spanning most of cosmic time, that are out of reach of current high-resolution spectroscopy. Spectropolarimetry enables detection and characterization of the magnetic fields (B-fields) and local environments of astrophysical objects. In addition, linear polarimetry will provide information on deviations from spherical symmetry, providing an extension of interferometry into a domain that is not restricted by the angular size of the objects but by their flux.

This aspect of POLLUX will be a very powerful tool in studies of contemporary stellar physics, addressing topics in star formation, stellar structure and evolution, and the nature of circumstellar environments, including light echoes of supernovae, which might be discernible in high redshift systems by the polarization across specific features in the spectrum. By providing linear and circular polarization (giving access to the four Stokes parameters) it will also probe the physics of accretion disks in many astrophysical situations, from supermassive black holes in AGNs to white dwarfs, detecting the polarized light reflected from Earth-like exoplanets or from their circumplanetary material (e.g. rings) and moons, and constrain the properties (sphericity) of stellar ejecta and explosions. The magnetospheric properties of planets in the solar system will be accessible in exquisite detail, while the influence of magnetic fields at the Galactic scale and in the IGM will also be within reach.

The parameter space opened by POLLUX is uncharted territory, so its potential for ground-breaking discoveries is high. It will also neatly complement and enrich some of the cases advanced for LUMOS[3], the multi-object spectrograph for LUVOIR.

# 2. SCIENCE OVERVIEW

## 2.1 Stellar magnetic fields across the Hertzsprung-Russell diagram

From the ground in the visible domain, we can obtain sophisticated tomographic mapping of structures on the surfaces of stars and their magnetic fields. However, to understand the formation and evolution of stars and their accompanying planets it is necessary to also explore their circumstellar environments. POLLUX will provide a powerful, high-resolution, full-Stokes (IQUV) spectropolarimetric capability to uniquely trace these structures. When combined with the immense light-gathering power of LUVOIR, it will deliver unprecedented views of the disks, winds, chromospheres and magnetospheres around a broad range of stars, as highlighted below (also see Figure 1).

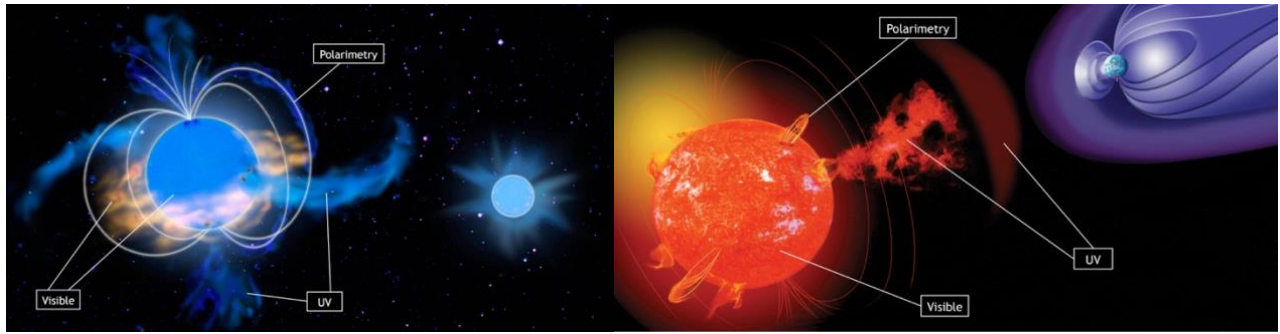

Figure 1: **Left**: Sketch of a hot star with its fossil magnetic field lines, channeled polar wind, surface spots, equatorial magnetosphere, corotating interaction regions, and a stellar companion. **Right**: Sketch of a cool star, with its dynamo magnetic field, surface faculaes and plages, wind, a coronal mass ejection, and a bow shock between the star and its planet. (© S. Cnudde)

Magnetic fields play a significant role in stellar evolution, while also being a key factor in both planet and star formation. We need POLLUX spectropolarimetry to identify and characterize the B-fields across all stellar masses, to address central questions such as:

- How do B-fields develop in the pre-main-sequence (PMS) phase? POLLUX will characterize the strong B-fields arising at the sheared interface between star and disk in the PMS phase (via MgII, FeII, and CII UV lines from the interface region), and will trace the flows via polarization measurements of the nearby continuum.

- How does the stellar dynamo build-up and evolve? As stellar rotation decouples from the young planetary disk, the B-field is predicted to get stronger and more complex[4]. POLLUX will investigate propagation of magnetic energy through the stellar atmosphere into the uppermost coronal layers and stellar wind.

- How do stellar dynamos form and evolve in cool stars? What is their impact on planet formation and evolution? POLLUX will reveal the magnetically mediated interaction between young cool stars and their planetary disks during dynamo formation and stabilization when planets and planetary atmospheres form.

- A small but significant fraction of massive stars have strong B-fields (~10%)[5]. However, very weak fields could be ubiquitous in massive stars. POLLUX will uniquely detect sub-Gauss B-fields in massive stars, providing a much-needed understanding of weak fields, field-decay mechanisms, and their impact on stellar evolution.

- POLLUX will enable studies of stellar B-fields beyond the Milky Way for the first time. These will open-up observations in the metal-poor Magellanic Clouds (MCs), probing the evolutionary processes linked to B-fields at metallicities equivalent to those at the peak of star-formation in the Universe.

POLLUX will also enable breakthroughs in other areas of contemporary stellar astrophysics, including:

- Jet formation: The mechanism driving the powerful jets from the interaction between disks around PMS stars and their magnetospheres is poorly constrained; POLLUX will transform our understanding of this dramatic part of star formation.

- Winds & Outflows: A large uncertainty in evolutionary models of massive stars is the link between rotation and mass lost via their winds, which limits our understanding of the progenitors of gamma-ray bursts, pair-instability SNe, and gravitational-wave sources. POLLUX observations in the Galaxy and MCs will map the density contrasts with stellar latitude required to calibrate these models.

- Novae & Supernovae (SNe): Both classical novae and SNe remnants provide excellent opportunities to learn about the production and lifecycle of dust. POLLUX spectropolarimetry is required to identify sites in the ejecta where dust condenses - allowing us to also infer the effects of kicks imparted by core-collapse SNe - and the shapes/sizes of the grains involved. Furthermore, polarization measurements of the unresolved ejecta during the earliest stages of the expansion will reveal its structural properties, especially sphericity, for novae and supernovae.

- White Dwarfs (WDs): A minority of WDs is strongly (>1 MG) magnetic[3], but little is known of the incidence of weaker fields. POLLUX UV spectropolarimetry will explore this regime for the first time to investigate the ubiquity of B-fields in WDs, as well as the processes generating their fields (e.g., amplification of fossil fields and/or binary mergers?)

A 100-hour POLLUX observing program would be transformational in our understanding of B-fields in stellar evolution. For instance, we could include UV monitoring of two low-mass sources (PMS T Tauri, evolved T Tauri) to directly characterize their B-fields and magnetic-energy transport, combined with observations of a first sample of tens of Galactic massive stars to assess the presence/strength of weak B-fields.

## 2.2 What are the characteristics of exoplanets atmospheres and how do planets interact with the host stars?

The characterization of exoplanets is key to our understanding of planets, including those in the Solar System. POLLUX unique, simultaneous, high-resolution and polarimetric capabilities in the UV are essential to unveiling the origins of the huge range of chemical and physical properties found in exoplanetary atmospheres[6] and to understand the interaction between planets and their host stars[7].

The line and continuum polarization state of starlight that is reflected by a planet is sensitive to the optical properties of the planetary atmosphere and surface and depends on the star-planet-observer phase angle (Figure 2).

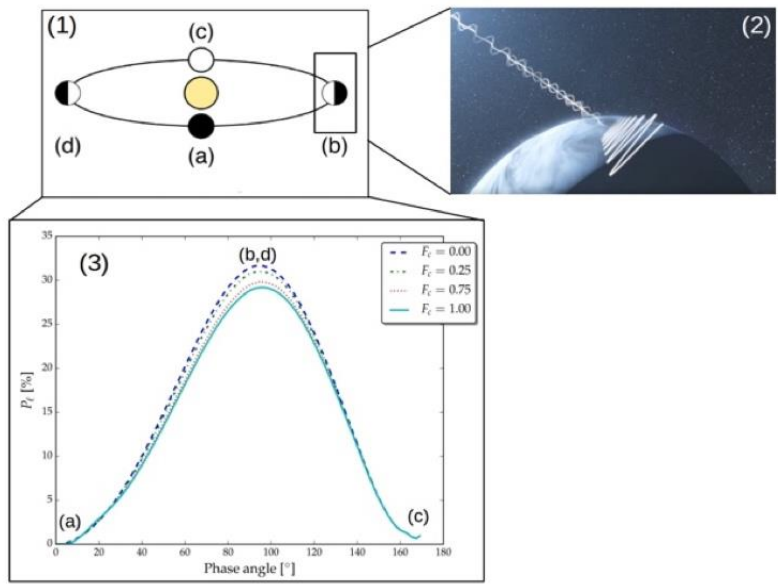

Figure 2. (1) & (2): Schematic of a planetary system in which the unpolarized stellar light becomes polarized through reflection by a planetary atmosphere. Panel 2 illustrates how the unpolarized stellar light coming from the top-left corner becomes linearly polarized following reflection by the planetary surface/atmosphere. Left (3): degree of polarization (in %) as a function of planetary orbital phase, labelled as in panel (1), at 300 nm (from Loic Rossi). The different lines indicate different atmospheric cloud coverages, where zero corresponds to the cloudless condition. POLLUX will measure the degree of polarization as a function of wavelength and planetary orbital phase.

Atmospheric gases are efficient scatterers at UV wavelengths and deviations from the expected polarization across the UV would reveal the presence and microphysical properties of aerosol and/ or cloud particles. This information is crucial for getting insight into a planet's climate. It would also strongly complement transit observations. POLLUX can significantly detect polarization signatures for close-in gas giants (and brown dwarfs) orbiting stars out to distances of 70 pc, which currently comprises over a dozen targets, two of them transiting.

Star-planet interactions (SPI) could generate detectable signatures in exoplanetary systems. Searching for SPI has developed into a very active field[8,9,10]. Evidence of SPI has been observed in UV stellar emission lines (e.g., NV[11]). POLLUX will be capable of studying the intensity of UV stellar emission lines forming in a wide range of temperatures as a function of planetary orbital phase, uniquely combining this information with measurements of the magnetic field derived from the Stokes V profiles of these lines. This will allow for the first time to identify the stellar regions mostly affected by SPI and hence to understand their origin.

At UV wavelengths, a hot Jupiter orbiting a Sun-like star with a two-day orbital period presents a maximum polarization of ≈30% (Figure 3), leading to a maximum polarization signal of the order of $7 \times 10^{-5}$, when observed unresolved from its star. With POLLUX mounted on a 15-m LUVOIR telescope and a binning size of 50 Å, the signal-to-noise ratio (S/N) necessary to detect such a signal can be obtained in less than 3 hours of shutter time for systems as far as 70 pc (Figure 3). Measuring and modeling the variation of the polarization signal as a function of the planet's orbital phase requires at least eight measurements spread across the orbit. A POLLUX 100-hour observing program would allow the detection and characterization of the UV polarimetric signature for over a dozen targets.

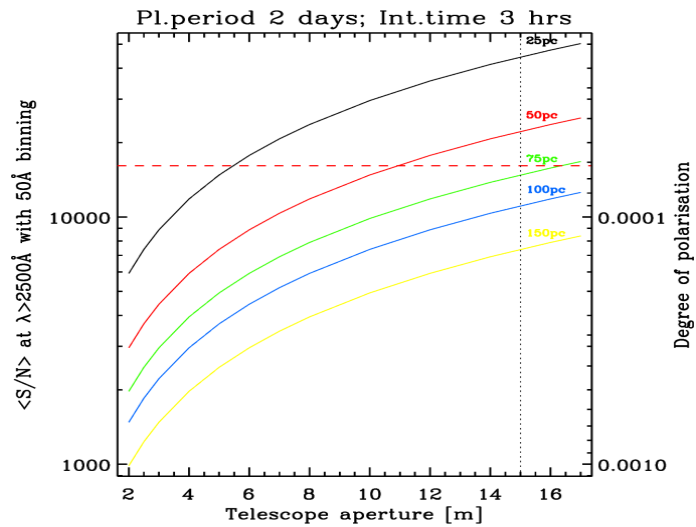

Figure 3: Average S/N in the near-UV obtained with 3 hours of POLLUX shutter time as a function of the distance of the target Sun-like star (line colour) and size of the telescope primary mirror. The red dashed horizontal line indicates the maximum near-UV polarization signal of a hot-Jupiter with a 2 days orbital period. The black dotted vertical line indicates the size of LUVOIR's primary mirror.

## 2.3 The various phases of the ISM and extragalactic IGM

Matter in the interstellar space is distributed in diverse, but well-defined phases that consist of (1) the hot, ionized ($T\sim10^{6-7}$ K) ISM that emits soft X-rays, (2) the warm neutral or ionized medium ($6000 < T < 10^4$ K), and (3) the cold ($T\sim10$–200 K) neutral medium and molecular star-forming clouds. Boundaries between the different phases play a fundamental role in the cooling of the gas and by consequence in Galactic evolution. However, it is not yet clear how these different phases trade matter and entropy. We need to address the multi-phase aspect of the ISM and the influence of different factors such as magnetic field, thermal/pressure (non-)equilibrium, metallicity, shocks, etc.

Gaining fundamental insights into the processes that influence the different phases requires the wealth of interstellar absorption features appearing in the UV spectra of hot stars and requires observing them at $R \geq 120{,}000$ to measure individual clouds and disentangle the different phases contributions. POLLUX will, for the first time, give access to all phases at the required high spectral resolution. It will enable simultaneous study of tracers of hot-gas (OVI, CIV), warm gas (OI, NI, and many singly-ionized species), as well as cold-HI gas like CI, and of the molecular phases through $H_2$ and CO lines. POLLUX will resolve the velocity profiles of the different $H_2$ rotational levels, yielding temperature and turbulence, and information on the formation of $H_2$ and its role as a coolant, e.g., in turbulence dissipation.

ISM studies also need to be generalized to external galaxies. POLLUX will enable tomography of chemical abundances in several hundred nearby galaxies using neutral gas absorption lines toward individual stars across the entire (stellar) body of galaxies or toward background quasars. Abundance variations across and between galaxies of different types and metallicities will document the dispersal/mixing spatial/time scales of newly produced elements and their nucleosynthetic origin, the role of infalling and outflowing gas in the metallicity build-up, and the dust production/composition through depletion patterns. Furthermore, cooling rates, physical conditions, and the molecular gas content can be determined spatially in the HI reservoir (with implications for the regulation of star formation and for galaxy evolution at large) and in neutral shells of HII regions (with implications for the star-formation process itself).

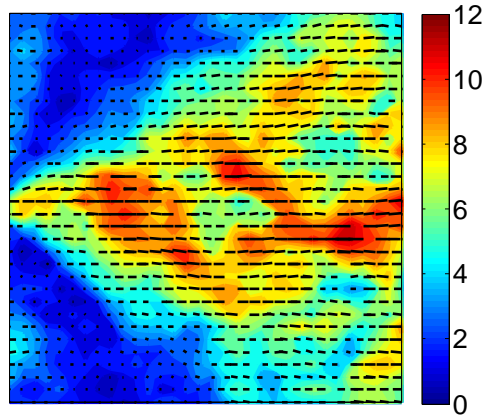

Figure 4: Synthetic polarization (%) map of the simulated diffuse ISM (1pc$^2$). Contour color: percentage of polarization in the SII 1250 Å absorption line. Orientation of the bars: polarization direction. Mean magnetic field =3 μG, oriented along the X-axis. This figure shows that the direction of polarization correctly indicates the magnetic field direction and that the expected polarization is mostly above 5%, within POLLUX capabilities.

A huge step forward in the knowledge of the Galactic magnetic field in terms of sensitivity, sky coverage and statistics has been made from recent dust polarization measurements. However, they tell us nothing on the distance of the magnetic field and its distribution in the different components or phases. The POLLUX spectropolarimeter will provide this information by its ability to detect linear polarization in UV absorption lines through Ground State Alignment[12]. It will open up access to the magnetic field 3-D, by measuring its orientation in different clouds, on limited distances along sight lines and on small scales (Figure 4). In a 100-hour program, observing several hundreds of hot stars with a S/N of 500 to measure linear polarization in optically thin absorption lines at a level of a few % will help understand the interplay of magnetic field and the different ISM phases. By discerning magnetic fields in gas with different dynamical properties, POLLUX will allow the first study of interstellar magnetic turbulence[13].

## 2.4 Dust composition and magnetic field strength of unresolvable regions of AGNs

POLLUX will offer unique insight into the still largely unknown physics of Active Galactic Nuclei (AGNs), which are believed to arise from accretion of matter by supermassive black holes in the central regions of galaxies. Some key signatures of accretion disks can be revealed only in polarized light, and with higher contrast at ultraviolet than at longer wavelengths. Specifically, high-resolution UV polarimetry will provide geometric, chemical and thermodynamic measurements of accretion disks in unprecedented detail. By probing the ubiquitous magnetic fields expected to align small, non-spherical dust grains on scales from the accretion disk out to the extended dust torus, POLLUX will be able to reveal the mechanisms structuring the multi-scale AGN medium.

The key information encoded into the polarized light will allow determinations of the mineralogy, structure and alignment of the smallest dust grains, together with on-line-of-sight magnetic-field strengths. Measurements of magnetic field strengths will also make constraints on the structure of magneto-hydrodynamic winds in nearby, broad absorption-lines systems accessible to the instrument. On larger scales, UV polarimetric studies of young star- forming regions will provide unprecedented insight into the enigmatic relation between onset of star formation and triggering of nuclear activity. Once activated, an AGN is expected to feed vast amounts of energy back into the interstellar medium of its host galaxy through radiation and shocks. This feedback can decrease or increase the star formation activity of the host, the physical conditions of which observational constraints from UV polarization with POLLUX will greatly help understand. In addition to the extended environment of AGNs, POLLUX will provide insight into the nature and lifetime of particles in relativistic jets, which are other key factors to fully understand AGN feedback on star formation in galaxies. By focusing on bright, low-redshift galaxies, it will be possible to obtain, for the first time, high-resolution spectra providing striking details on the structure and physics of AGNs. Complemented with the optical and infrared instruments on board LUVOIR, POLLUX will constitute a groundbreaking means of assessing the important role played by AGNs on galaxy evolution (Figure 5).

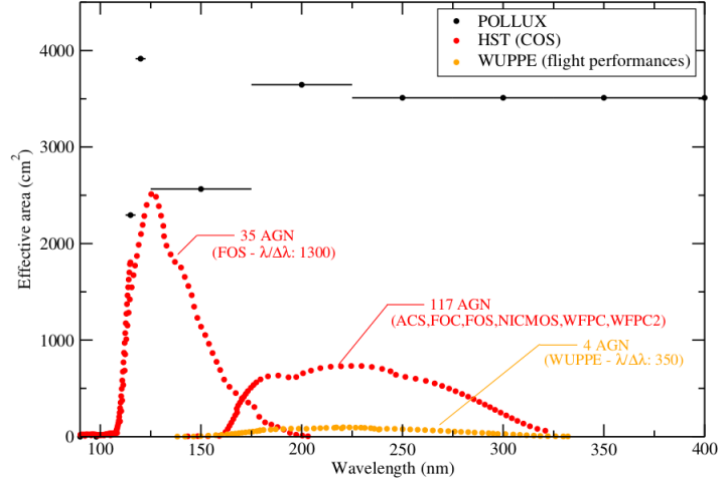

Figure 5: Effective area of POLLUX (in black; assuming 135 m² for LUVOIR with a RC telescope), compared to those of previous space-based UV spectrographs, as indicated. Only 4 AGN were observed with WUPPE. A total of 117 were observed with different polarimetric instruments on board HST, out of which only 35 in the UV with HST/FOS (for which an effective area similar to that of HST/COS was adopted). POLLUX will enable observations of hundreds of AGN over a wide UV spectral domain, and with much larger spectral resolution than ever achieved.

## 2.5 Testing fundamental physics and cosmology using absorption lines towards quasars

Physics and cosmology as we know today have been remarkably successful in reproducing most of the available observations with only a small number of parameters. However, it also requires that 96% of mass-energy content of the Universe is in mysterious forms (dark energy and dark matter) that has never been seen in the laboratory. This shows that our canonical theories of gravitation and particle physics may be incomplete, if not incorrect. Improving the sensitivity of current observational constraints is therefore of utmost importance, irrespective of whether it is consistent with the current standard physics—in which case it will reject other scenarios—or whether it will instead favor new physics.

Absorption-line systems produced by intervening gas in the spectra of background sources provide original sensitive probes of fundamental physics and cosmology. The high UV spectral resolution of POLLUX on LUVOIR will open a unique window on such probes, in particular (1) the measurement of the primordial abundance of deuterium, (2) the stability of fundamental constants over time and space and (3) the redshift evolution of the cosmic microwave background (CMB) temperature. The D/H ratio can be estimated from the DI and HI Lyman series lines. Any change in the proton-to-electron mass ratio ($m=m_p/m_e$) translates into a relative wavelength change of the $H_2$ Lyman and Werner lines. Finally, the CMB radiation excites the CO molecules so that the relative population in different rotational levels (measured through the electronic bands in the UV) is an excellent thermometer for the CMB temperature.

We remark that, while these are independent probes, they are intimately related by the underneath physics. For example, models involving varying scalar- photon couplings[14] also affect the temperature-redshift relation so that constraining this relation is complementary to a search for varying fundamental constants. The Big Bang nucleosynthesis calculations of the D/H ratio are also dependent on the fundamental constants and can be altered if new physics is at play[15].

The baseline specifications of LUVOIR/POLLUX are S/N~100 per resolution element in 1h for $F_\lambda=10^{-14}$ erg s$^{-1}$ cm$^{-2}$ Å$^{-1}$. This means that for known quasars with low-z $H_2$ absorbers as observed by HST/ COS[14], we can reach a precision of a $\Delta\mu/\mu$ ~ a few $10^{-7}$ in 20–30h (Figure 6). This is about an order of magnitude better than the current best limits in less observing time than typical used to achieve those (around 5 x $10^{-6}$ with UVES on the Very Large Telescope). Similarly, the achieved precision on TCMB scales directly with the S/N ratio (0.1 K at S/N~100 and R~100,000 when current limits are DT~1K). The same is again true with the D/H ratio, although a very high resolution is less critical since the corresponding lines are typically thermally broadened above 10 km s$^{-1}$.

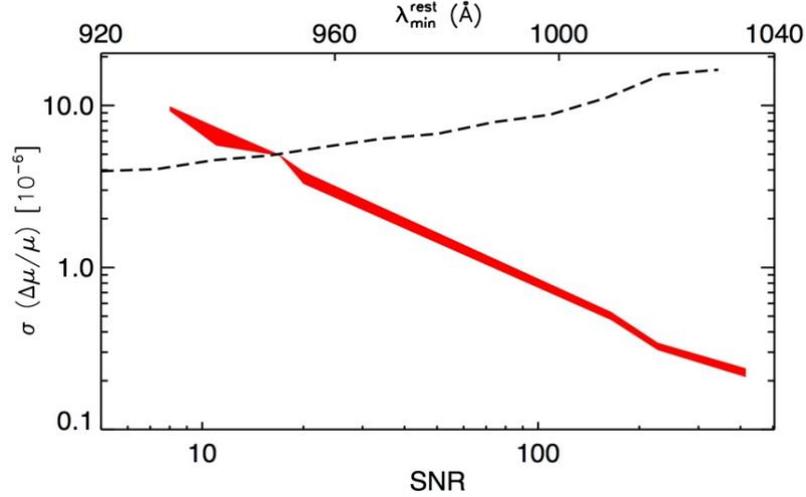

Figure 6: Expected error on $\Delta\mu/\mu$ as a function of signal-to noise ratio (red band, abscissa scale) and minimum wavelength (dashed line, absorber's rest-frame, top axis) covered by a R=120,000 spectrum.

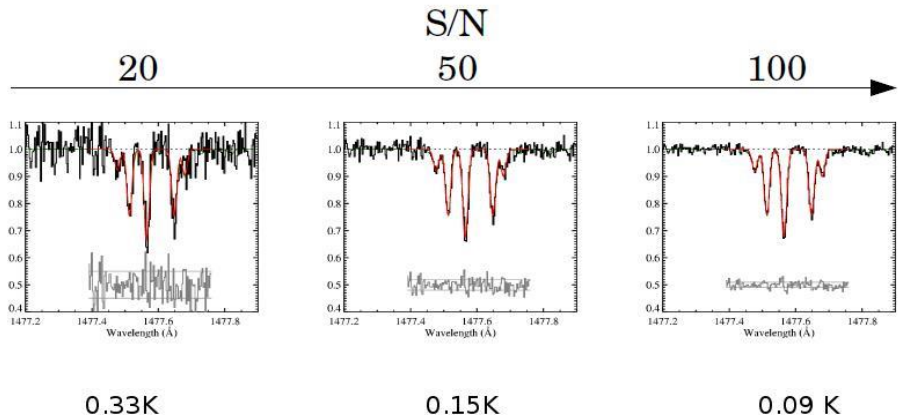

Figure 7: Simulated CO electronic absorption profile at R=120,000 and different S/N ratios. The error on the measured excitation temperature is indicated below each panel.

## 2.6 Solar system: surfaces, dust scattering and auroral emissions

UV observations uniquely probe the surface of telluric bodies of the solar system. They diagnose their volcanic and plume activity, their interaction with the solar wind and their composition in the frame of space weather and exobiology/habitability fields. Very few UV polarimetric observations were obtained so far with WUPPE[16], revealing in particular the Io surface as spatially covered by 25% $SO_2$ frost with polarization variations associated to different volcanic regions. POLLUX will primarily characterize volcanism and/or plume activity of icy moons from polarized solar continuum reflected light and spectral UV albedo. Its high sensitivity is necessary to track any organic and ice composition of the crust of comets and Kuiper Belt objects from their UV spectrum.

The giant planets' UV aurorae are mainly radiated from H and $H_2$ atmospheric species, collisionally excited by accelerated charged particles precipitating along the auroral magnetic field lines. Aurorae thus directly probe complex interactions between the ionosphere, the magnetosphere, the moons and the solar wind. Precipitation of auroral particles is additionally a major source of atmospheric heating, whose knowledge is needed to assess the energy budget, the dynamics and the chemical balance of the atmosphere. The narrow Field-of-View of POLLUX will measure the bright complex aurorae of Jupiter and Saturn, the fainter ones of Uranus and catch those of Neptune, only seen by Voyager 2[17]. The high spectral resolution will be used to finely map the energy of precipitating electrons from partial spectral absorption of $H_2$ by hydrocarbons[18,19] and the thermospheric wind shear from the H Ly-α line (Figure 8, left panels[20]).

The asymmetric profile of the H Ly-α line in the Jovian auroral region (Figure 8, left panels) should be resolved by POLLUX: for an exposure time of 10 minutes, and considering the brightness of the wings of the Ly-α line measured by HST/STIS, the two wings should produce 68±8 cts/px and 34±6 cts/px respectively, implying a velocity larger than 4 km/s which can be measured precisely by POLLUX.

The albedo of a region of 100 $km^2$ can be derived with an accuracy of 0.1% with a spectral resolution of 1 nm with a 5-min exposure only near 320 nm. The linear polarization of Io surface varies between 1 and -10% between 220 and 320 nm; near 320 nm, the linear polarized signal observed by POLLUX for a binning size of 1 nm in 5 minutes should have a S/N~10, which is better than WUPPE (Figure 8, right panels).

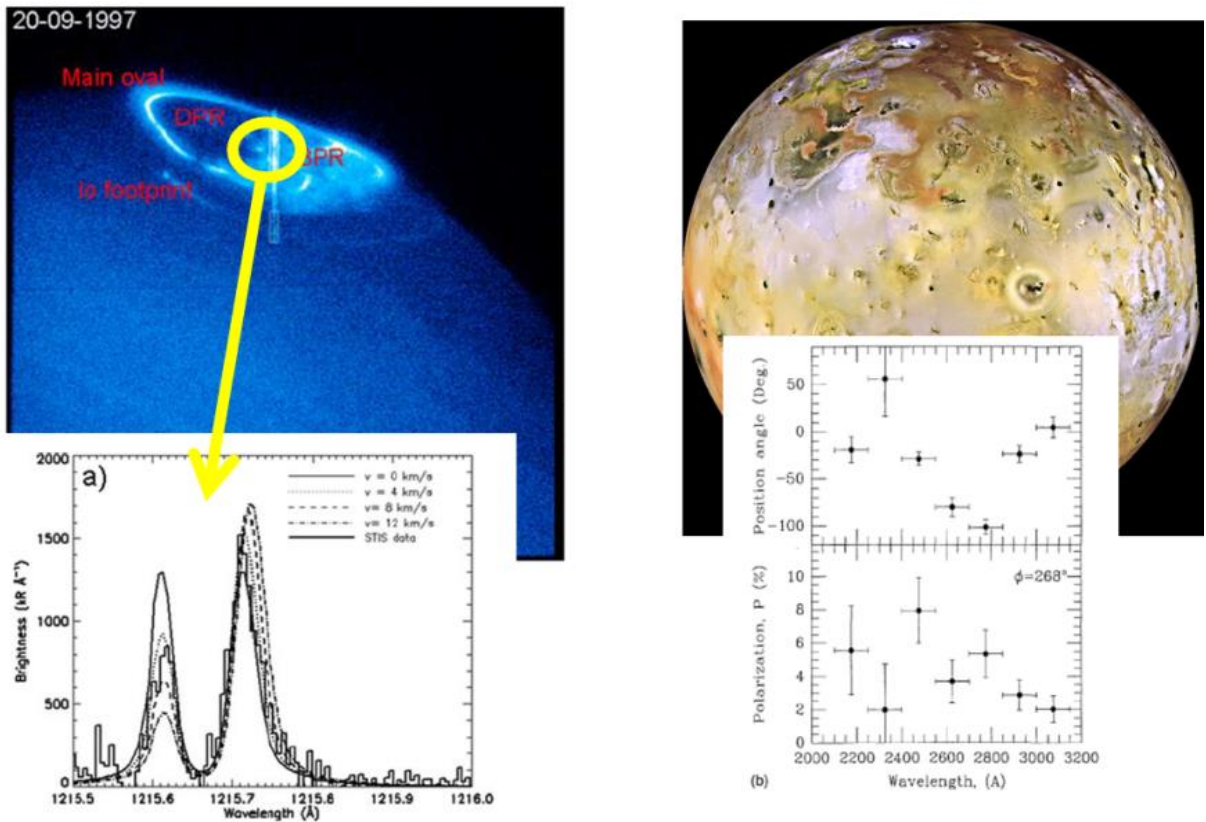

Figure 8. Left: auroral emissions observed on Jupiter by HST. The Lyα emission spectrum in the polar auroral region was obtained with STIS. Its strong asymmetry can be reproduced with a wind shear reaching 4-8 km/s[16]. Right: Io observed by Galileo (©Galileo Project, JPL, NASA). The linear polarisation of the surface measured by WUPPE between 220–320 nm presents complex variations with wavelength and provides information on the volcanic activity of the surface of Io[16].

# 3. HIGH-LEVEL REQUIREMENTS

The science goals described in the previous section lead to the technical requirements for POLLUX presented in the Table 1 below. Based on these requirements, a design concept for POLLUX instrument has been proposed[2]. In short:

- The spectral range will be split between 3 channels – far-UV (90-124.5 nm), medium-UV (118.5-195 nm) and near-UV (195-390 nm), essentially because it allows to use dedicated/optimised optical elements, coatings and detectors and polarimeter for each band, hence obtain a gain in efficiency. It also allows us to maintain a reasonable size for each channel and respective components, while maintaining the required resolution.
- Each channel is equipped with its own dedicated polarimeter followed by a high-resolution spectrograph.
- MUV+NUV channels are recorded simultaneously, while the FUV is recorded separately (temporal separation).
- The FUV and MUV boundaries are set relative to the Lyman-α line, such that this line is always present on both channels. The MUV and NUV boundary is set such that a maximum of one full octave falls in each channel (here the NUV) but this may be changed in a later design.
- The instrument will allow either spectropolarimetric, or pure spectroscopic modes.
- POLLUX can be fed by the light coming from the telescope or from sources in the calibration unit.

Table 1: High-level requirements derived from the science cases, with comments in the last column

| Parameter | Requirement | Goal | Reasons for requirement |
|---|---|---|---|
| **Wavelength range** | 97 - 390 nm | 90 – 650 nm | 97nm to reach Lyγ line. 390 nm to reach CN line at 388 nm in comets |
| **Spectral resolving power** | 120,000 | 200,000 | Resolve line profiles for ISM, Solar System and Cosmology science cases |
| **Spectral length of the order** | 6 nm | ≥6 nm | To avoid having broad spectral lines spread over multiple orders |
| **Polarisation mode** | Circular+linear (= IQUV) | | |
| **Polarisation precision** | $10^{-6}$ | Detect polarization of hot Jupiters | |
| **Aperture size** | 0.03" | 0.01" | Avoid contamination by background stars in Local Group galaxies |
| **Observing modes** | spectropolarimetry and pure spectroscopy | | |
| **Radial velocity stability** | Absolute = 1 km/s and relative = 1/10 pixel | | Avoid spurious polarization signature |
| **Flux stability** | 0.1% | | Probe flux and polarization correlation in WDs |
| **Limiting Magnitude** | V=17 | | To reach individual stars in MCs |
| **Calibration** | Dark, bias, flat-field, polarization and wavelength calibration | + Flux calibration | |

# 4. SUMMARY

In this paper, we presented an overview of the wide variety of science that could be addressed with POLLUX onboard LUVOIR. This instrument has unique capabilities, as it will provide high-resolution (R ≥ 120,000) spectropolarimetry across the UV bandpass (90-390nm). Such capabilities are essential for determining the physical, chemical and/or magnetic properties of astrophysical objects. They will help bring decisive clues to questions at the core of the program for the LUVOIR observatory, concerning the origin and evolution of the Universe and the life within it.

## ACKNOWLEDGEMENTS

The authors acknowledge contributions from the entire LUVOIR and POLLUX consortia. We would like to especially thank Kevin France (Univ. of Colorado – Boulder), Matthew Bolcar (NASA GSFC) and Aki Roberge (NASA GSFC). We also want to acknowledge the crucial contribution of Eduard Muslimov (LAM), Arturo Lopez Ariste (IRAP) and Maelle Le Gal to the optical design of POLLUX and its polarimeters.